\documentclass[12pt]{iopart}
\usepackage{iopams}
\usepackage{graphicx}
\usepackage{subfigure}

\begin{document}

\setcounter{subfigure}{0}
\setcounter{footnote}{0}

\title{A high-precision rf trap with minimized micromotion for an In${}^+$ multiple-ion clock}

\author{K Pyka, N Herschbach, J Keller and T E Mehlst\"aubler}

\address{Physikalisch-Technische Bundesanstalt, Bundesallee 100, D-38116 Braunschweig, Germany}
\eads{\mailto{karsten.pyka@ptb.de}, \mailto{tanja.mehlstaeubler@ptb.de}}

\begin{abstract}

We present an experiment to characterize our new
linear ion trap designed for the operation of a many-ion optical
clock using $\mathrm{^{115}In^+}$ as clock ions. For the
characterization  of the trap as well as the sympathetic cooling
of the clock ions we use $\mathrm{^{172}Yb^+}$. The trap design
has been derived from finite element method (FEM) calculations and
a first prototype based on glass-reinforced thermoset laminates
was built. This paper details on the trap manufacturing process
and micromotion measurement. Excess micromotion is measured using photon-correlation spectroscopy with a resolution of $\mathrm{1.1\,nm}$ in motional amplitude, and
residual axial rf fields in this trap are compared to FEM calculations.
With this method, we demonstrate a sensitivity to
systematic clock shifts due to excess micromotion of $|(\Delta\nu/\nu)_\mathrm{mm}|=8.5\times10^{-20}$. Based on the measurement of axial rf fields of our trap, we estimate a number
of twelve ions that can be stored per trapping segment and used
as an optical frequency standard with a fractional inaccuracy of
$\leq1\times10^{-18}$ due to micromotion.

\end{abstract}

\submitto{\NJP}

\maketitle

\tableofcontents

\section{Introduction}\label{s_introduction}

The improvement of optical clocks based on single-ion spectroscopy
has lead to frequency comparisons with a relative inaccuracy of
$|\Delta\nu/\nu|<1\times10^{-17}$ in the last
years~\cite{Chou_2010}. Most promising candidates have shown to be
two-species systems, in which the clock ion is sympathetically
cooled and the detection is realized through quantum logic
spectroscopy~\cite{Schmidt_2005}. To reduce systematic shifts
related to excess micromotion for such a two-ion system, linear
Paul traps are used, ideally having zero axial micromotion due to
their geometry.

While single-ion clocks show an excellent potential to reach
lowest frequency inaccuracy, they are limited in their short-term
stability by the intrinsically low signal-to-noise ratio (SNR), as
only one ion contributes to the clock signal. One approach to
improve the short-term stability of the clock measurement, thus
reducing lengthy averaging times, is to increase the spectroscopy
pulse time on long-lived atomic states~\cite{Chou_2010,
Roberts_1997}. This naturally limits the number of available clock
ion candidates and places severe requirements on the clock laser
stability.

Proposals have been made to develop an optical frequency standard
based on many ions with increased SNR
\cite{Herschbach_2011, Champenois_2010}. In our approach,
we sympathetically cool indium ions with ytterbium ions. Here, the
direct spectroscopy of indium as the clock ion avoids quantum
logic and facilitates scaling-up the clock read-out of many
ions~\cite{Herschbach_2011}.

However, with a larger number of clock ions the residual axial
micromotion of a linear ion trap and its associated systematic
frequency shifts becomes an issue. In~\cite{Chou_2010}, for
example, clock ion and cooling ion are separated by a distance of
$3\mathrm{\,\mu m}$ along the trap axis. A relative frequency
shift due to axial micromotion of
$(\Delta\nu/\nu)_\mathrm{mm}\approx-2.7\times10^{-17}$ is observed, when the
ions swap their position. For an optical clock with many ions, it
is a primary goal to engineer trap structures capable of trapping
linear chains of ions with lowest possible axial rf fields and a
high-level control of the ion dynamics.

A great effort in scalable trap engineering has been already made
in quantum information experiments with trapped ions
(\cite{Madsen_2004,Home_2006,Stick_2006,Schulz_2006, Hensinger_2006,Leibrandt_2009,Tanaka_2009,Pearson_2006,Allcock_2011}).
Here, fast ion transport is required and relevant time scales are
in the millisecond regime or below, thus electrode dimensions are
comparatively small. In~\cite{Herschbach_2011} we proposed a
design for a scalable linear ion trap for an optical clock, in
which up to ten ions can be stored per trapping segment within the
Lamb-Dicke regime. Since observed excess heating rates scale with
the ion-electrode distance $d$ like $\Delta E_\mathrm{heat}/\rmd t\propto
d^{-4}$~\cite{Turchette_2000,Daniilidis_2011}, the trap electrode dimensions were
chosen to be of millimeter size, to reach long interrogation
times, unlimited by the heating of the ion. A strong effort is put
on optical access and precise dimensions of the trap segments to
minimize axial rf fields to a level required for ultra-precise
clock spectroscopy.

In this paper we present our new experimental setup for testing
high-precision ion traps and demonstrate successful operation of a
first prototype. The atomic system used to characterize the ion
trap is $\mathrm{^{172}Yb^+}$. The laser system for this ion, the
vacuum system and the detection scheme are described in
section~\ref{s_experimental_setup}. In section~\ref{s_trap}, the
design and the fabrication process of the prototype trap are
explained in detail. In the last section, we focus on
high-resolution micromotion measurements and compare
experimentally determined axial rf fields to FEM calculations of
this trap. We demonstrate a sensitivity of photon-correlation
spectroscopy to micromotion induced frequency shifts of
$|(\Delta\nu/\nu)_\mathrm{mm}| < 10^{-19}$. With this method, we
evaluate our ion trap to be capable of trapping up to 12 ions
along the trap axis for high-precision spectroscopy with a
micromotion induced relative frequency inaccuracy below
$1\times10^{-18}$.

\section{Experimental setup}\label{s_experimental_setup}

\subsection{Laser systems for the trapping of $\mathrm{^{172}Yb^+}$}\label{ss_lasers}

\begin{figure}[hbtp]
\flushleft
\hspace{2.5cm}
\includegraphics[width=0.83\textwidth]{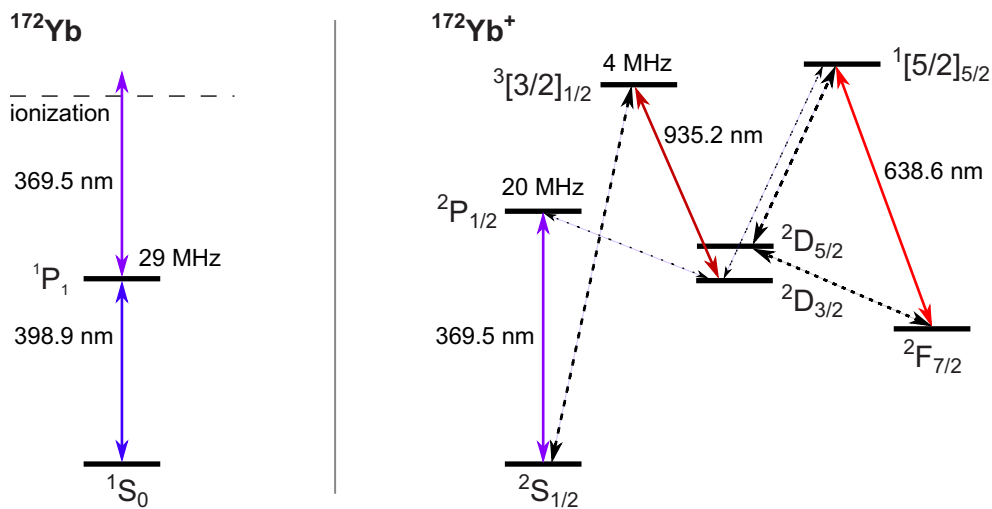}
\caption{Left: Photoionization of $\mathrm{^{172}Yb}$ via two-step
laser excitation. Right: Partial level-scheme of
$\mathrm{^{172}Yb^+}$ with atomic transitions used for cooling and
repumping.}\label{yb_scheme}
\end{figure}

The advantage of the even isotope $\mathrm{^{172}Yb^+}$, used to
characterize the ion trap, is the absence of a hyperfine structure
and thus a low number of electronic states that have to be
addressed. Four different diode lasers are used for
photoionization, cooling and detection, and repumping. An overview
of the relevant energy levels including their linewidth and
transition wavelength is given in figure~\ref{yb_scheme}.

For photoionization, a frequency-doubled external cavity diode
laser (ECDL) system is used to resonantly excite neutral
$^{172}$Yb to its $\mathrm{^1P_{1}}$ state. A single pass through
a Brewster-cut periodically-poled potassium titanyl phosphate
(PPKTP) crystal is sufficient to generate a power of
$P_\mathrm{399nm}=80\mathrm{\,\mu W}$, out of which $7\,\mathrm{\mu W}$ are sent to
the atoms. The beam is guided through the trap vertically,
whereas the atomic beam of neutral Yb passes horizontally through
the trap. Thereby, Doppler-shifts and broadening due to the high
temperature of the atoms are avoided. In fact, a scan of the laser
shows a full-width half maximum of only $35\,\mathrm{MHz}$ at a
saturation parameter of $s = 2\Omega^2/\Gamma^2 \lesssim 1$, where $\Omega$ is the Rabi-frequency and $\Gamma$ the decay rate of the excited state.

An optically amplified and frequency-doubled diode laser system
(Toptica TA SHG pro)  delivers an output power of
$P_\mathrm{369nm}=55\,\mathrm{mW}$ at a wavelength of
$369.5\,\mathrm{nm}$. This light is used to cool and detect the
ions by driving the $\mathrm{^2S_{1/2}\ \leftrightarrow \
^2P_{1/2}}$ transition of $\mathrm{^{172}Yb^+}$. It is also used
for the second excitation step of the photoionization process. The
laser is divided into three beams, all sent to the trap via
polarization-maintaining optical fibers. All beams are collimated
and imaged into the trap with a waist of
$\omega_\mathrm{369nm}=80\,\mathrm{\mu m}$, where the cooling beam in
the vertical direction is overlapped with the photoionization
beam, see figure~\ref{trap_laser_detection_scheme}. The power used
for Doppler cooling and detection of a single ion is $\approx 3\,\mathrm{\mu W}$, which gives a saturation parameter of $s_\mathrm{369nm}=0.6$ for each beam. In order to cool and
perform spectroscopy on chains of ions in more than one trap
segment simultaneously, it is necessary to expand the beams along
the trap axis over several millimeters. This will be done in the
near future and will make use of the high power provided by this
laser.

Due to the decay channel from the cooling transition to the
long-lived $\mathrm{^2D_{3/2}}$ state, constant cooling of the ion
is only possible by depleting the population of this dark state.
Therefore, another laser beam with a power of $20\,\mathrm{mW}$ is
sent to the ions, driving the $\mathrm{^2D_{3/2} \leftrightarrow
\, ^3[3/2]_{1/2}}$ transition at a wavelength of
$935.2\,\mathrm{nm}$. The beam is guided in a single-mode
polarization-maintaining optical fiber and imaged with a
$f=500\,\mathrm{mm}$ lens
 along the trap axis (Z, see figure~\ref{trap_laser_detection_scheme}), having a minimum waist of
$\omega_\mathrm{935nm}=125\,\mathrm{\mu m}$ at the trap center.

Collisions with the background gas in the vacuum chamber can
populate the metastable $\mathrm{^2F_{7/2}}$ state by a
combination of non-radiative and radiative
decays~\cite{Schauer_2009}. This state has an estimated lifetime
of several years~\cite{ Roberts_1997,Fawcett_1991}, so the ion is
lost for the experimental sequence and the decoupling from the
cooling-cycle leads to heating and eventual loss from the trap.
Also, quenching of the metastable state population via collisions
with heavier background molecules such as $\mathrm{H_2O}$ is observed. For
a deterministic clean-out of the $\mathrm{^2F_{7/2}}$ state, a
second repump laser at $638.6\,\mathrm{nm}$ is used in the
experiment. About $5.5\,\mathrm{mW}$ of this light is overlapped with the
$935.2\,\mathrm{nm}$ laser beam, again having a waist of
$\omega_\mathrm{639nm}=125\,\mathrm{\mu m}$ at the trap center.

\begin{figure}[hbtp]
\flushleft
\hspace{2.5cm}
\includegraphics[width=0.6\textwidth]{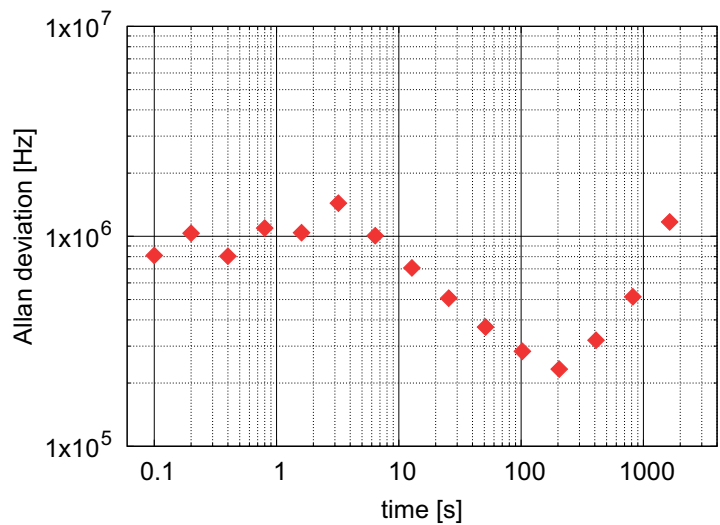}
\caption{Temporal stability of the wavemeter read-out. Shown is
the Allan deviation of a typical frequency measurement of our
stable reference laser at $822\,\mathrm{nm}$, with sub-Hz
linewidth and drift of $< 1\,\mathrm{Hz/s}$, using the WS-7
wavemeter.}\label{wm_adev}
\end{figure}

A wavelength meter based on a Fizeau interferometer\footnote{High
Finesse, WS-7} with an 8-channel multi-mode fiber switch serves as
a frequency reference for all lasers. The specified absolute
frequency accuracy is $60\,\mathrm{MHz}$ in a range of
$350\,\mathrm{nm}\textrm{ to }1100\,\mathrm{nm}$ for a single-mode fiber
connection, and $200\,\mathrm{MHz}$ for the multi-channel switch
operation. Slow frequency drifts of the diode lasers are
compensated with an electronic feedback on their piezo actuators
by our experiment control software. The reproducibility of the
measured frequencies is about $1\,\mathrm{MHz}$ from shot to shot. With a
temperature stability of $\sigma_\mathrm{T}=0.2\,\mathrm{K/h}$ a drift of less
than $3\,\mathrm{MHz}$ per hour is observed. Figure~\ref{wm_adev} shows the
stability of the wavemeter read-out against an ultra-stable laser
with a sub-Hz linewidth and a few $100\,\mathrm{mHz/s}$ drift.

\subsection{Experimental apparatus and detection scheme}\label{ss_vacuum_system}

The vacuum system is designed for testing and characterization of
different trap geometries and provides versatile optical and
electrical access. The cylindrical chamber is based on a DN250
flange with adapter flanges on the side and bottom for windows and
electrical feedthroughs.

\begin{figure}[hbtp]
\flushleft
\hspace{2.5cm}
\includegraphics[width=0.83\textwidth]{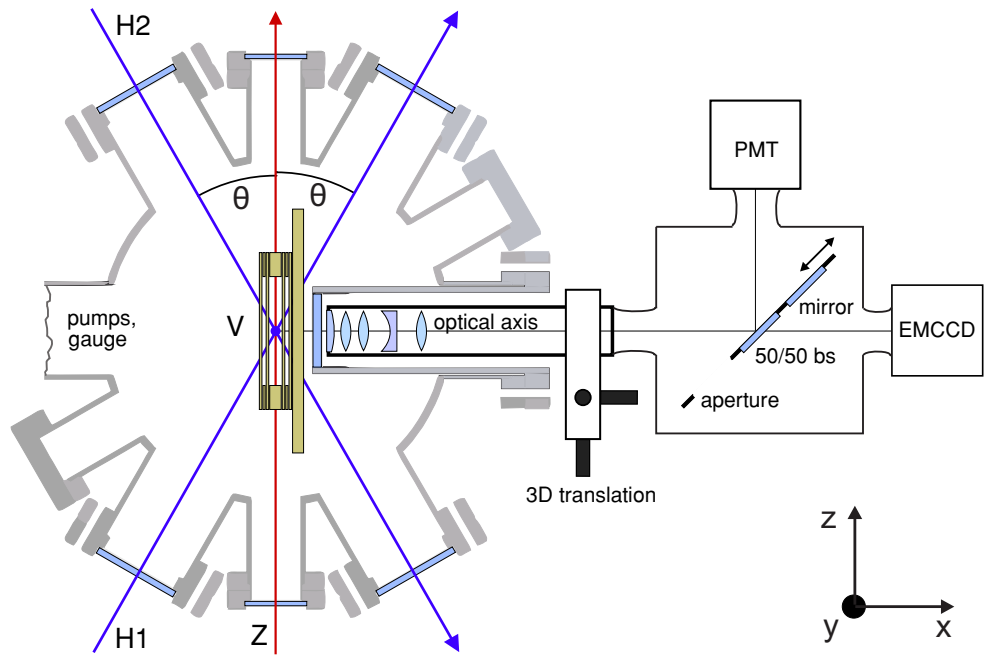}
\caption{Schematic top view of the experimental setup. The trap is
placed in front of the re-entrant vacuum viewport. Three laser
beams (\textbf{H1, H2, V}) at $369.5\,\mathrm{nm}$ are used for
the measurement of micromotion in all spatial dimensions. Laser
beams along \textbf{Z} provide light at $638.5\,\mathrm{nm}$ and
$935.2\,\mathrm{nm}$ for repumping. The photoionization laser at
$398.9\,\mathrm{nm}$ is overlapped with the beam in
y direction.}\label{trap_laser_detection_scheme}
\end{figure}

Six windows are mounted pairwise in the horizontal plane and allow
laser beams to pass along the trap axis (Z) as well as under an
angle of $\theta=\pm 25\,^{\circ}$ to the z-axis (H1 and
H2). One vertical laser beam (V) passes perpendicularly to the
horizontal plane through the ion trap. With this setup, we are
able to measure excess micromotion of the ion in all three
dimensions and characterize the trap. A detailed scheme is shown
in figure~\ref{trap_laser_detection_scheme}.

For detection, a re-entrant viewport is mounted horizontally and
perpendicularly to the trap axis with a distance of
$23\,\mathrm{mm}$ between the vacuum window and the trap center. A self-built lens system with standard spherical 1 inch lenses and a working distance of $37\,\mathrm{mm}$ provides a light collection efficiency of
$2\,\%$. The lens design is a retrofocus lens with a numerical aperture $N\!A=0.27$. In order to assure wavefront errors smaller than $\lambda/4$ the lens positions have been optimized with a commercial ray-tracing software. With a magnification $V=25$ and a CCD-chip pixel size of $16\,\mathrm{\mu m}$, we obtain a spatial resolution of $0.6\,\mathrm{\mu m/px}$ on our camera.

The detection scheme is twofold: it is possible to either detect
with the electron multiplying CCD (EMCCD) camera\footnote{Andor:
DU-897, $\mathrm{512\,px\times512\,px}$, quantum efficiency
$Q\!E=35\,\%$ for $\mathrm{200\,nm-370\,nm}$} or with a
photomultiplier\footnote{Hamamatsu: R7207-01, Bialkali window, $Q\!E\geq20\,\%$ for $\mathrm{160\,nm-650\,nm}$} (PMT). Also, it
is possible to detect ions with both devices simultaneously using
a 50/50 beam splitter. The latter configuration was used during
the measurements presented in this paper.

The vacuum in the apparatus is maintained by an ion getter pump,
having a nominal pumping speed of $20\,\mathrm{mbar\,l/s}$ for air
and a titanium sublimation pump in a DN63 tube, with a calculated
pumping speed of $490\,\mathrm{mbar\,l/s}$ for hydrogen. The
residual pressure after pre-baking, re-opening and inserting the
ion trap, followed by a modest bake-out at $60\,\mathrm{^\circ C}$ is
$1.0 \times 10^{-8}\,\mathrm{Pa}$.

\begin{figure}[hbtp]
\flushleft
\hspace{2.5cm}
\includegraphics[width=0.6\textwidth]{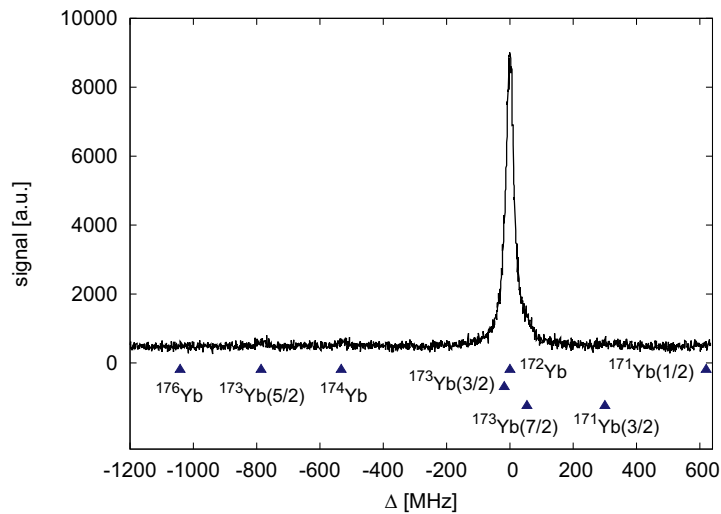}
\caption{$399\,\mathrm{nm}$ laser scan with isotope enriched
$\mathrm{^{172}Yb}$ sample. The expected resonances of the various
isotopes~\cite{Das_2005} are indicated by
triangles.}\label{yb_oven_scan}
\end{figure}

An 8-pole feedthrough provides a high current connection to the Yb
and In ovens inside the vacuum chamber. The oven design consists
of a vertical tantalum tube with an outer diameter of
$1.02\,\mathrm{mm}$ and an inner diameter of $0.86\,\mathrm{mm}$.
On both ends, copper wires are spot-welded to provide the heating
current. A $0.4\,\mathrm{mm}$ diameter hole on the side of the
tantalum tube generates an atomic beam. To collimate the
atomic beams of both ovens, a copper shield with two slits of
$\approx1\,\mathrm{mm}$ in the horizontal direction is placed
between the ovens and the trap, assuring atomic flux through the
loading segment, while the spectroscopy segments do not get
contaminated. An isotope-enriched sample of ytterbium is used, as
the natural abundance of the isotopes $\mathrm{^{172}Yb}$ and
$\mathrm{^{173}Yb}$ are $22\,\%$ and $16\,\%$,
respectively. Their photoionization resonances are separated by
only $55\,\mathrm{MHz}$ and show considerable overlap. A frequency
scan of the photoionization laser is shown in
figure~\ref{yb_oven_scan}.

A 41-pole feedthrough provides the connection for the various dc
voltages for axial confinement of the ions and for micromotion
compensation in our segmented linear Paul trap. The dc voltages
are generated by a 13 bit analog output PCI-card\footnote{National
Instruments: NI 6723 with 32 analog outputs}. A self-built
low-pass filter and a distribution box combine the axial
confinement voltages and radial compensation voltages that are
applied on a common trap electrode. A detailed scheme is shown in
figure~\ref{trap_electrode_scheme}. For the electrodes with
strongest passage, the compensation voltages provided by the
PCI-card are divided down to improve the resolution of the field,
which is applied to shift the ions in the radial direction
($U_\mathrm{tc}$). The range and resolution of all three fields are
summed up in table~\ref{Udc_table}.

\begin{figure}[hbtp]
\flushleft
\hspace{2.5cm}
\includegraphics[width=0.6\textwidth]{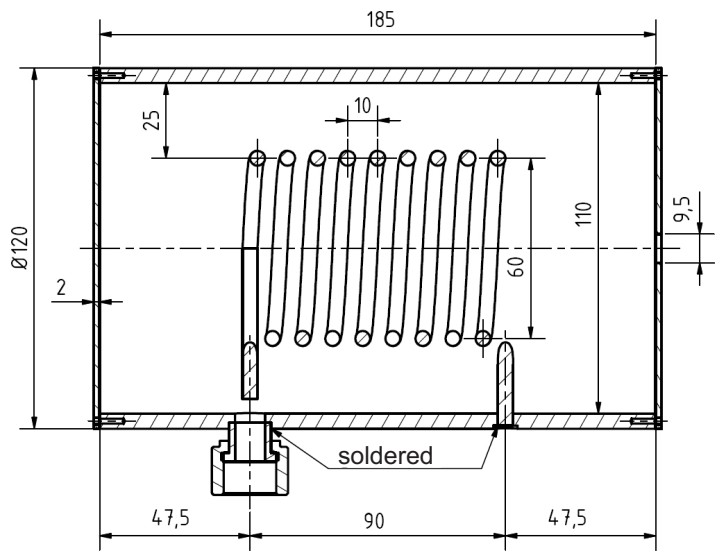}
\caption{Schematic drawing of the helical resonator for rf voltage
generation in our setup. All dimensions are given in
mm. The resonator coil is made of 5 mm thick Cu wire. The circuit is fed from the right side 
by a primary coil with a single winding, not shown in the drawing. Connected to the trap load of $16\,\mathrm{pF}$, we achieve a quality factor of $Q_\mathrm{loaded}=640$.}\label{resonator}
\end{figure}

Furthermore, an HN type coaxial feedthrough connects the trap with
the helical resonator, which generates the high-voltage rf field
for the radial confinement, that is necessary in order to achieve high secular frequencies for $\mathrm{^{172}Yb}^+$.
To reduce heating of the system due to power dissipation, the resonator is designed for a high quality factor according to~\cite{Macalpine_1959}. With a free-standing coil, made of a $5\,\mathrm{mm}$ thick copper wire, an unloaded quality factor
of $Q_\mathrm{unloaded}=1050$ has been realized. Connected to
the trap, the quality factor is $Q_\mathrm{loaded}=640$ and the
resonance frequency for driving the trap is
$\Omega_\mathrm{rf}=2\pi\times25.67\,\mathrm{MHz}$. The amplitude of the
rf voltage is measured by a calibrated thin wire pick-up, which is
placed inside the helical resonator tank. A schematic drawing of the resonator is given in
figure~\ref{resonator}.

With this system, the ion trap can be driven with an rf
voltage amplitude of $1500\,\mathrm{V}$ using an rf power
$<2\,\mathrm{W}$.

\section{The ion trap}\label{s_trap}

\subsection{Specifications}\label{ss_specifications}

\begin{figure}[hbtp]
\flushleft
\hspace{2.5cm}
\includegraphics[width=0.6\textwidth]{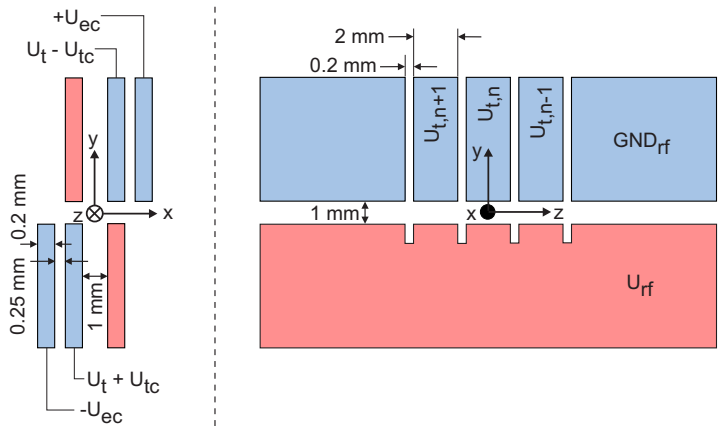}
\caption{Trap geometry and electronic configuration. All rf ground
electrode segments are dc isolated from each other. With
individual voltages $U_\mathrm{t,n}$ axial confinement is
realized. A differential voltage $U_\mathrm{tc,n}$ provides
compensation fields in radially diagonal direction in each segment
$n$. A differential voltage $U_\mathrm{ec,n}$ on the outer
compensation electrodes provides an independent second field
vector to move the ions to any position in the xy-plane.}\label{trap_electrode_scheme}
\end{figure}

The trap introduced in this paper is based on a design published
in~\cite{Herschbach_2011}. The main goal is to obtain a scalable
structure of an array of linear Paul traps, where chains of about
10 ions can be stored in the Lamb-Dicke regime in each trapping
segment, with axial rf fields at negligible level. This means
micromotion induced systematic fractional frequency shifts need to
be below $1\times10^{-18}$. For this, FEM calculations
had been carried out to estimate the effects of alignment uncertainties
and machining tolerances on the rf potential. In order to test the
trap design and the new experimental setup, a simplified prototype
trap has been built from easily machinable thermoset wafers,
featuring one loading segment and two spectroscopy segments. The
realized electrode structure with five electronically isolated
segments and the geometric dimensions are shown in
figure~\ref{trap_electrode_scheme}.

The rf electrodes $U_\mathrm{rf}$ only carry the high rf voltage
for the radial confinement of the ions, whereas the rf ground
electrodes $G\!N\!D_\mathrm{rf}$ provide dc voltages for the axial
confinement of the ions as well as the micromotion compensation in
all three dimensions. The inner electrodes opposite to the rf
electrodes are used for axial confinement as well as micromotion
compensation, while the outer electrodes are used for micromotion
compensation only. Table~\ref{Udc_table} shows the dc voltages
that can be applied, together with their resolution and the
corresponding dc electric field calculated at the position of the
ion.

\begin{table}
\caption{DC voltages that can be applied to the trap electrodes.
The corresponding maximum electric field at the position of the
ions and its resolution is given along the specified trap
axis.}\label{Udc_table}
\begin{indented}
\item[]\begin{tabular}{@{}llllll} \br
& trap axis & $\Delta U_\mathrm{dc}$ & $U_\mathrm{dc,range}$ & $\Delta E_\mathrm{dc}$ & $E_\mathrm{dc,range}$\\
\mr
$U_\mathrm{tc}$    & x & $\mathrm{0.6\,mV}$ & $\mathrm{\pm\lineup\0 2.5\,V}$ & $\mathrm{0.32\,V/m}$ & $\mathrm{1325\,V/m}$\\
    & y &  &  & $\mathrm{0.40\,V/m}$ & $\mathrm{1650\,V/m}$\\
$U_\mathrm{ec}$    & x & $\mathrm{2.9\,mV}$ & $\mathrm{\pm12\,V}$ & $\mathrm{0.45\,V/m}$ & $\mathrm{1860\,V/m}$\\
    & y &  &  & $\mathrm{0.05\,V/m}$ & $\mathrm{\lineup\0 204\,V/m}$\\
$U_\mathrm{t}$     & z & $\mathrm{2.9\,mV}$ & $\mathrm{\pm12\,V}$ & $\mathrm{0.30\,V/m}$ & $\mathrm{\lineup\0 612\,V/m}$\\
\br
\end{tabular}
\end{indented}
\end{table}

For the axial confinement, $U_\mathrm{t}$ is applied to the
electrodes neighbouring the used trap segment. A typical value for
most measurements presented in this paper is $\omega_\mathrm{ax,Yb} =
2\pi\times116\,\mathrm{kHz}$ at $U_\mathrm{t,n-1} = U_\mathrm{t,n+1} = 4\,\mathrm{V}$
and $U_\mathrm{t,n}=0\,\mathrm{V}$. As those electrodes are controlled
independently, they can also be used for micromotion compensation
as well as shifting the ion along the trap axis to measure
residual rf electric fields. Here, a field resolution of
$\Delta E_\mathrm{z,min}=0.30\,\mathrm{V/m}$ corresponds to a spatial
shift of $0.3\,\mathrm{\mu m}$ at $\omega_\mathrm{ax,Yb} =
2\pi\times116\,\mathrm{kHz}$.

For micromotion compensation in radial direction,
$U_\mathrm{ec}$ and $U_\mathrm{tc}$ are used.
$U_\mathrm{ec}$ generates a field, which has its strongest
vector component ($90\,\%$) along the x-axis and is solely used to
shift the ion along this direction. $U_\mathrm{tc}$ generates a
field with strong components in both x and y direction, which
makes it necessary to compensate with $U_\mathrm{ec}$ in x
direction when shifting the ion along the y-axis. The equally
spaced quadrupole electrode geometry leads to an efficient rf
trapping potential in radial direction with a calculated loss
factor of $L=1.29$~\cite{Schrama_1993}.

For clock operation, radial secular frequencies of more than $1\,\mathrm{MHz}$
can be reached with indium ions at an rf trap voltage of $1500\,\mathrm{V}$. For
the trap characterization with $\mathrm{^{172}Yb^+}$, a lower rf
voltage amplitude of $U_\mathrm{rf}= 810\,\mathrm{V}$ is used, resulting
in radial secular frequencies of $\omega_\mathrm{rad,Yb} =
2\pi\times 484\,\mathrm{kHz}$. At this rf amplitude the trap depth is
$2.2\,\mathrm{eV}$ in x direction and $3.4\,\mathrm{eV}$ in y
direction. The principal axes of the trap are rotated with respect
to the x-axis by about $36\,^\circ$ and
$126\,^\circ$.

\subsection{Trap fabrication}\label{ss_trap_fabrication}

The prototype ion trap is made of Rogers4350B$\texttrademark$, a glass-reinforced,
ceramic-filled hydrocarbon thermoset with low rf losses
($\tan \delta < 0.0037$)\footnote{see
http://www.rogerscorp.com, not available anymore, but current 4360
shows same specifications}. This makes it possible to generate a
high trapping rf voltage with low input power. The electrodes are
made of laser structured $200\,\mathrm{\mu m}$ thick wafers, with
a $35\,\mathrm{\mu m}$ conductive copper layer and a
$10\,\mathrm{\mu m}$ gold thick-film. A thin nickel layer is used
as an adhesion promoter.

\begin{figure}[hbtp]
\flushleft
\hspace{2.5cm}
\subfigure{}{\includegraphics[width=0.47\textwidth]{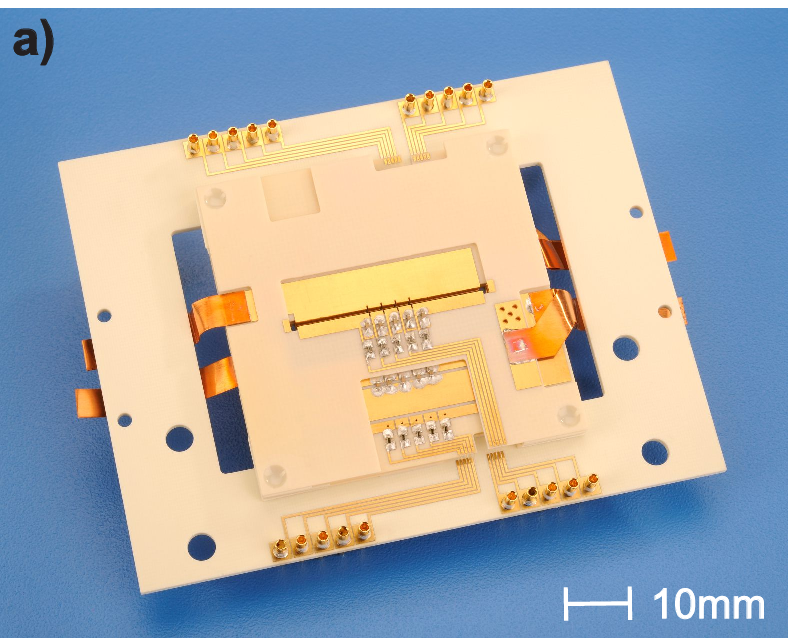}}
\hspace{0.5cm}
\subfigure{}{\includegraphics[width=0.3\textwidth]{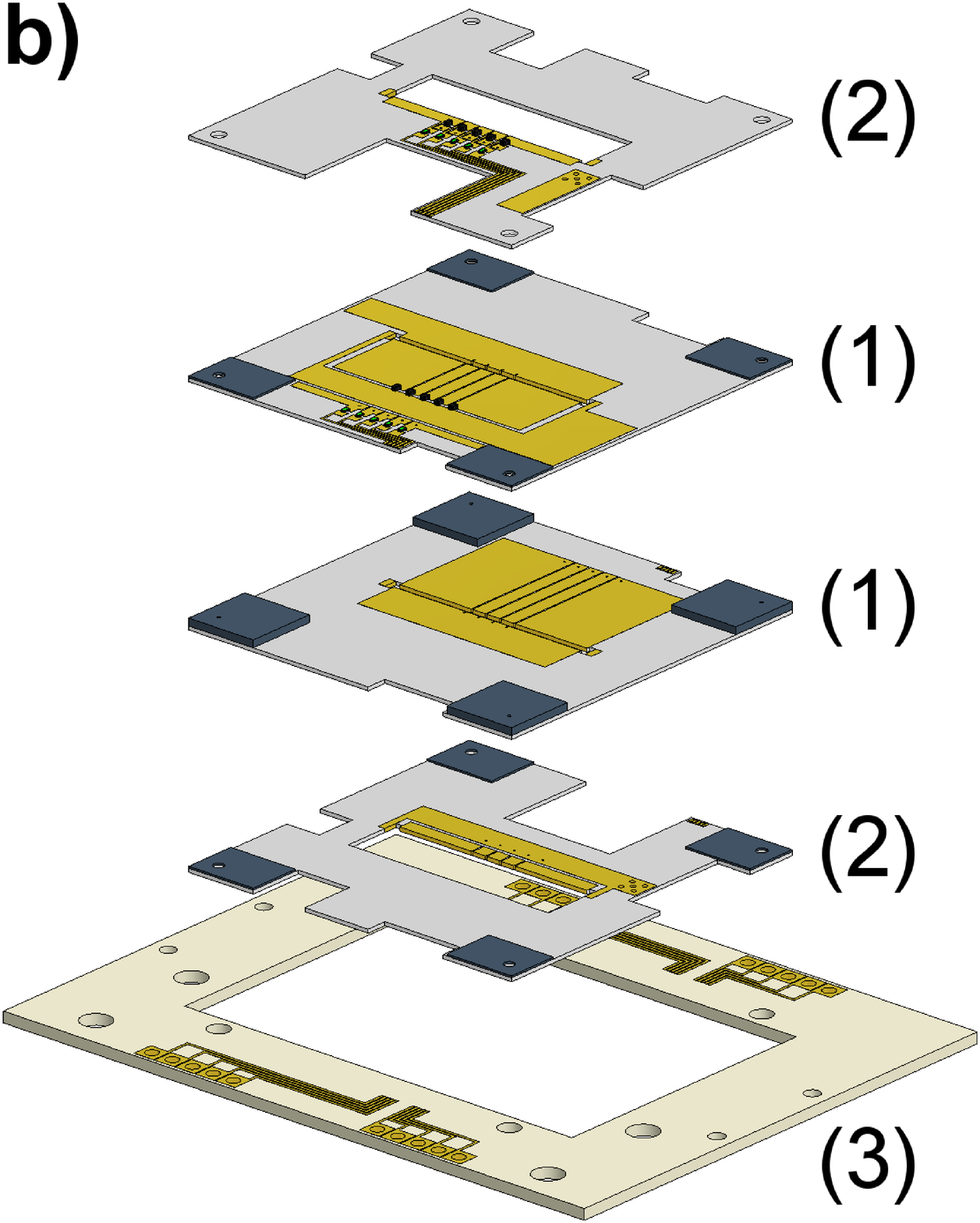}}
\caption{a) Photograph of the assembled trap stack with
on-board filter electronics, connector pins for dc voltages on the
carrier board and rf feed copper strips. b) Scheme of the
trap assembly with two trap (1) and two compensation (2) boards
with spacers (dark grey) on the corners to provide optical access
for the laser beams.}\label{trap_scenic_view}
\end{figure}

Four boards are stacked on top of each other and mounted on a
carrier board as shown in figure~\ref{trap_scenic_view}.
Two identical boards (1), one of which is rotated by $180\,^\circ$
around the trap axis relative to the other, form the quadrupole
trap. They are separated by four  $1\,\mathrm{mm}$ thick spacers
placed in the corners. An additional board (2) is attached on top
of each of the trap boards, with spacers of $0.25\,\mathrm{mm}$
thickness. The stack is glued on a $1.5\,\mathrm{mm}$ thick
carrier board (3), which provides gold pins as dc voltage
connectors to the multipin feedthrough and a mount for the rf
connection. The rf lead is a $0.1\,\mathrm{mm}$ thick, oxygen-free
copper foil cut into $5\,\mathrm{mm}$ wide strips of equal length
to avoid phase shifts on the rf electrodes.

To prevent coupling of rf power into the dc voltage sources as
well as coupling of high-frequency noise onto the trap electrodes,
low-pass filters are mounted directly on the trap boards close to
the dc electrodes. Non-magnetic and UHV-proof SMD
components\footnote{Resistors: $R=300\,\mathrm{k\Omega}$, Barry
Industries - partnumber: RP0402BA-3003 JN-91. Capacitors:
$C=4.7\,\mathrm{nF}$, Novacap - partnumber: 0402 C472 J500 PH-HB.
Low-pass cutoff frequency: $\nu_\mathrm{cutoff}=113\,\mathrm{Hz}$} are
soldered with a lead free UHV compatible solder\footnote{Kester:
80Sn19Ag1Cu}. The copper foil strips used for the rf connection
are soldered onto the boards as well.

After fitting all boards with the electrical components, they were
aligned under the cross-hair of a microscope\footnote{Leitz UWM}.
Translation stages with a $1\,\mathrm{\mu m}$ scale and a rotation
stage with a $1\,\mathrm{arc\,second}$ scale provided high-precision
alignment control. The boards were fixed one after the other using
a glue with low outgassing and low shrinkage\footnote{Optocast
3410 Gen2, see http://outgassing.nasa.gov/ for vacuum
compatibility}. This glue has the advantage of being both heat and
UV curable. After stabilizing the boards with a broadband UV lamp,
the complete stack was heated in an oven for $30\,\mathrm{min}
\textrm{ at }130\,\mathrm{^{\circ}C}$ to ensure that all glue was cured.
The dc leads on the individual electrode boards were then
connected to the carrier board with a ball bonder using
$30\,\mathrm{\mu m}$ thick gold wires. Finally, the gold pins were
soldered to the carrier board for the dc voltage connection. The
link to the vacuum feedthrough is made with Kapton coated wires
with crimped connectors.

As a first check, the magnetization of the single components of
the trap was measured with a fluxgate in a $\mu$-metal box with a
sensitivity of $\leq100\,\mathrm{nT}$. No effect could be measured
for either the SMD parts or the trap boards. Apparently, the thin
nickel layer used as an adhesion promoter does not show bulk
behavior yet and therefore no ferromagnetic properties. Only for
the gold-plated pins for the dc voltage connection a residual
magnetization of about $1\,\mathrm{\mu T}$ at a
probe distance of $7\,\mathrm{mm}$ could be measured. Scanning the probe over the
trap stack, field amplitudes of up to $2\,\mathrm{\mu T}$ were
measured near the pins at the edge of the carrier board, but no
field was resolved in the vicinity of the trap slit. For the final
trap design, the gold pins will be replaced by soldered dc
connections on the carrier board.

\section{Characterization of the prototype trap}\label{s_characterization}

\subsection{Trap operation}\label{ss_trap_operation}

\begin{figure}[hbtp]
\flushleft
\hspace{2.5cm}
\includegraphics[width=0.6\textwidth]{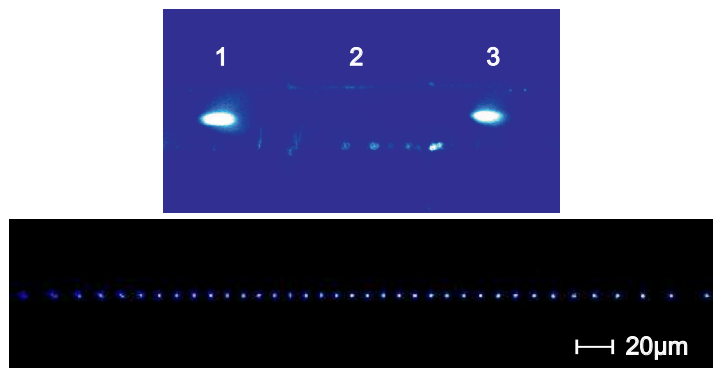}\label{ion_chain}
\caption{Top: trapping and shuttling of $^{172}\mathrm{Yb}^+$ ions
in ion trap, illuminated by beams of the cooling laser along H1
and H2. The stray light of the laser beams makes the segmented trap
structure visible. Bottom: picture of a linear Coulomb crystal of
37 $^{172}\mathrm{Yb}^+$ ions, trapped in segment 3 and imaged
with the self-built diffraction limited lens.}\label{ion_shuttle}
\end{figure}

As a first test of the ovens and the ion trap, a single lens was
used to observe the whole trapping region. The fluorescence of
neutral Yb atoms was observed in the trace of the photoionization
beam passing through the loading segment, segment 3 in
figure~\ref{ion_shuttle}. By moving the laser beam through all trap
segments and detecting atoms only in segment 3, we were able to
verify the alignment of the atomic beam aperture. Ions were loaded
and shuttled into all three trap segments. With our self-built
detection lens the magnification was then increased to
$V=25$, which leads to a resolution of $\approx
0.6\,\mathrm{\mu m/px}$ and makes it possible to resolve single ions in a
Coulomb crystal in one trap segment, see figure~\ref{ion_shuttle}.
Deterministic loading of single ions was demonstrated.

The radial secular frequencies were measured for a single
$^{172}\mathrm{Yb}^+$ ion by amplitude modulation of the rf
electric field. For an rf voltage amplitude of
$U_\mathrm{rf}=810\,\mathrm{V}$ and a dc voltage
$U_\mathrm{t}=0.05\,\mathrm{V}$, two radial secular frequencies of
$\omega_\mathrm{rad,1}=2\pi\times 490\,\mathrm{kHz}$ and
$\omega_\mathrm{rad,2}=2\pi\times 472\,\mathrm{kHz}$ were measured. Here,
the rf voltage is deduced from the calibrated pick-up voltage
$U_\mathrm{mon}$ of the helical resonator with
$K=U_\mathrm{rf}/U_\mathrm{mon}=5400\pm 340$.
The observed trap frequencies are, within a few percent, in good
agreement with the calculated rf potential.

\subsection{Photon-correlation spectroscopy}\label{ss_spectroscopy}

A single $\mathrm{^{172}Yb^+}$ ion is loaded into the trap to
measure residual micromotion in our trap by photon-correlation
spectroscopy. This technique measures the rf field induced motion
of the ion via its 1st order Doppler-shift on a broad atomic line
with linewidth $\Gamma$. In order to resolve the weak modulation
of the ion fluorescence at the trap frequency, its scattering rate
is correlated with the phase of the rf trap voltage. A
time-to-amplitude-converter (TAC) generates pulses with a height
dependent on the time $T$ between a START-signal, triggered by the
detection of a photon by the PMT, and a STOP-signal generated by
the rf voltage of the trap. A multi-channel-analyzer yields a
histogram of these pulses sorted in height and thus, in time
difference $T$. Thereby, the modulation amplitude of the ion's
fluorescence at the trap frequency $\Omega_\mathrm{rf}$ is
observed. A detailed description of the method can be found
in~\cite{Berkeland_1998}.

\begin{figure}[hbtp]
\flushleft
\hspace{2.5cm}
\subfigure{}{\includegraphics[width=0.4\textwidth,angle=0]{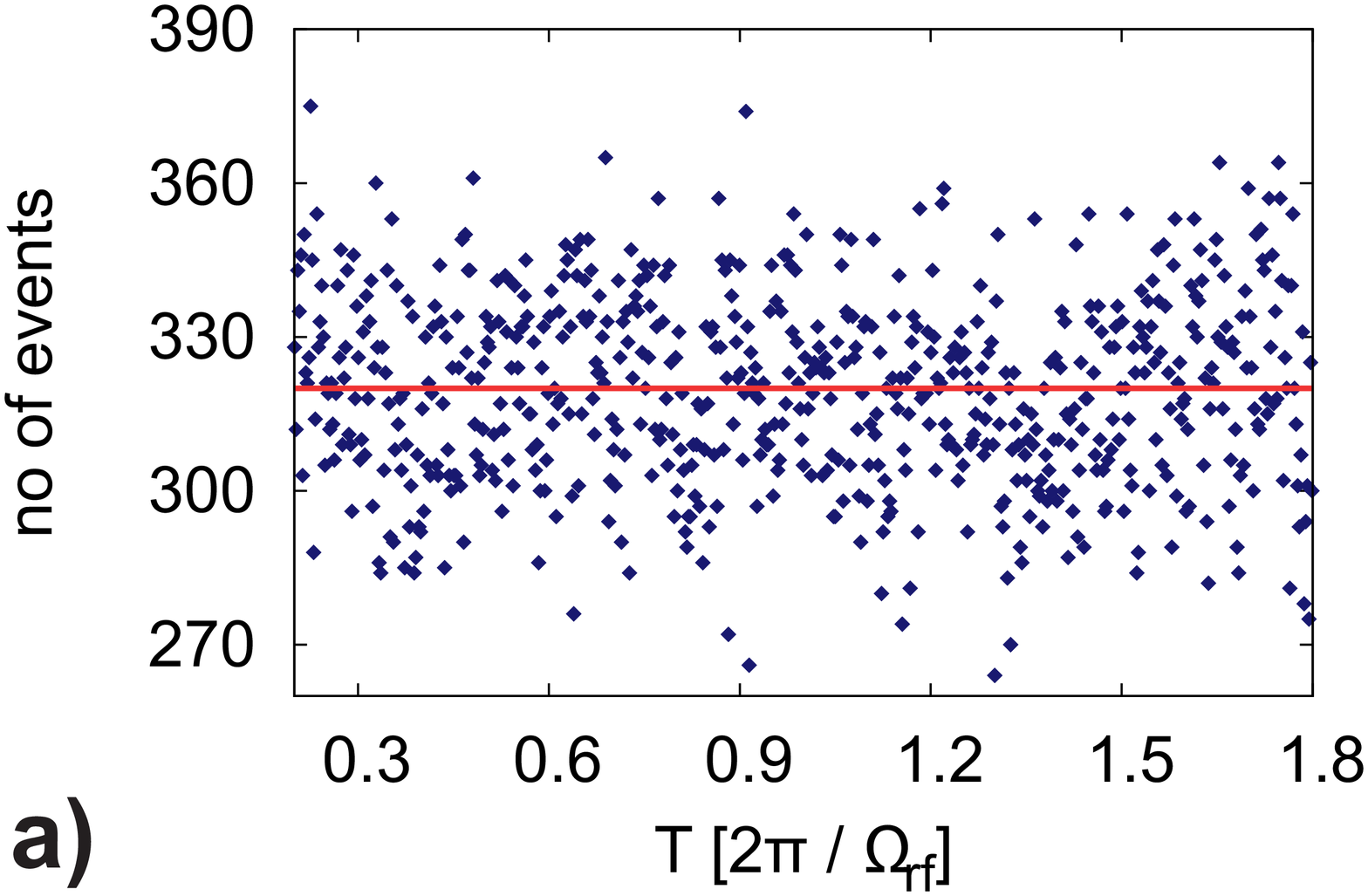}}\hspace{0.3cm}
\subfigure{}{\includegraphics[width=0.4\textwidth,angle=0]{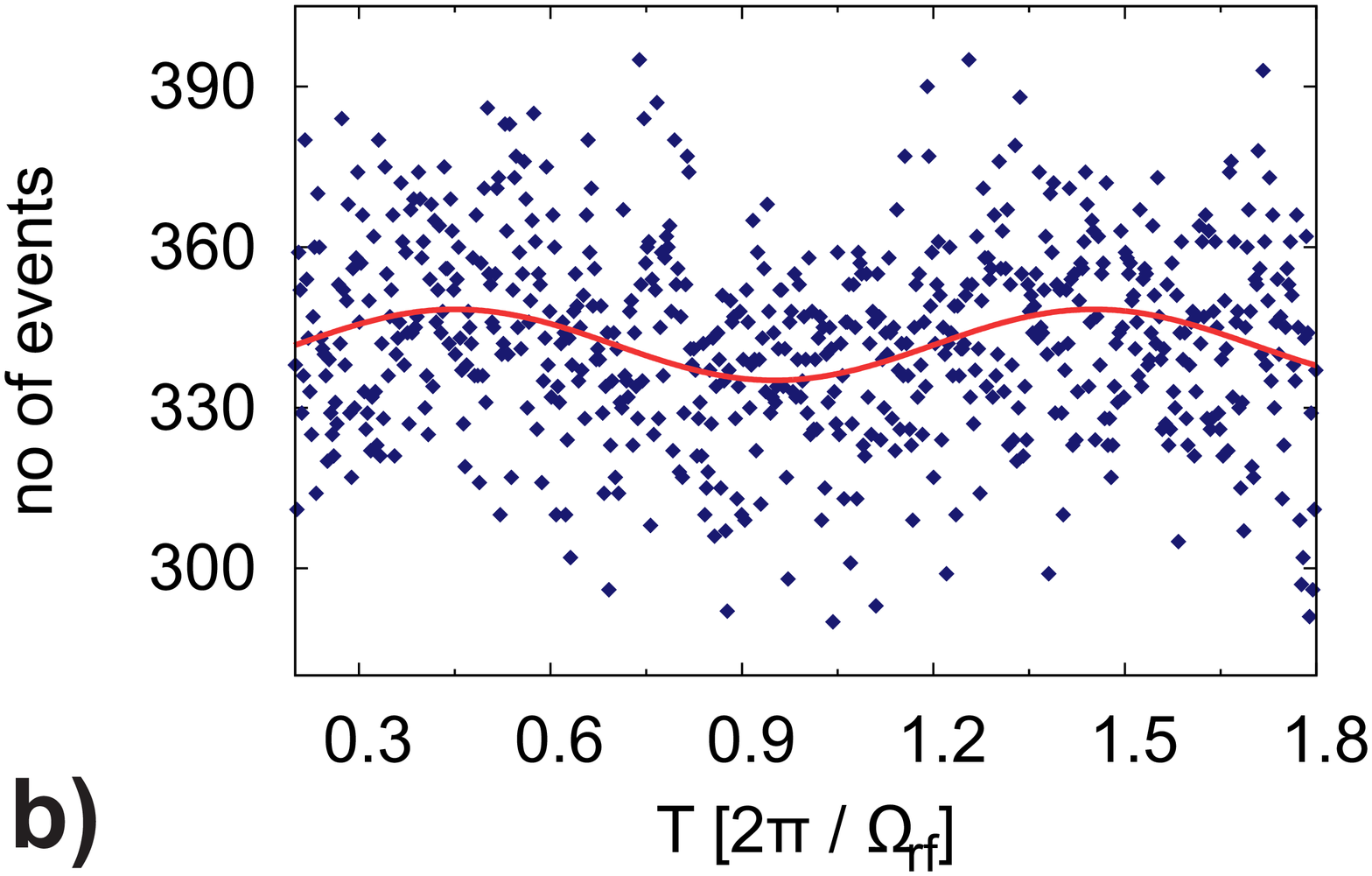}}
\caption{Photon-correlation signal of micromotion along y
direction as measured with vertical laser beam (V). The
acquisition time is $30\,\mathrm{s}$. a) Ion with optimized
micromotion. Restricting the phase to $-\pi/2 < \varphi_\mathrm{mm} < \pi/2$ the fit yields $S_\mathrm{mod}/S_\mathrm{max} = 0\pm 0.0017$. b) Ion shifted by $\Delta
U_\mathrm{ec}= 5.8\,\mathrm{mV}$ (corresponding to $\Delta x=57\,\mathrm{nm}$) in x
direction. The red line is a sine fit to the data with
$S_\mathrm{mod}/S_\mathrm{max} = 0.011$.}\label{mm_resolution}
\end{figure}

As an example, figure~\ref{mm_resolution} shows the histograms of
two measurements of the radial micromotion component along the
y-axis of a single ion as measured by the laser beam along V.
For each measurement, the data acquisition time is
$30\,\mathrm{s}$. In the measurement shown in
figure~\ref{mm_resolution}, where micromotion is detected along the
vertical laser beam, a weak laser beam along H1 is present to
cool the ion in the axial direction. The average photon count rate
at a detuning of $\omega_\mathrm{laser}-\omega_\mathrm{ion}=-\Gamma/2$
is about $10\,800\,\mathrm{s^{-1}}$, including a background of
about $1800\,\mathrm{s^{-1}}$.

Using the laser beams in H1, H2 and V, respectively, the
micromotion can be measured in all dimensions. The velocity
amplitude of rf induced ion motion $v_\mathrm{mm}$ is
evaluated using the linearization of the line profile, which is a
valid approximation for $k\,v_\mathrm{mm}\ll\Gamma$. In this
case, the signal contribution $S_\mathrm{det,i}$ of each beam
$i$ can be written as
\begin{eqnarray}\label{eq_signal_fit}
S_\mathrm{det,i} & = & \frac{S_\mathrm{max,i}}{2} + S_\mathrm{mod,i}\sin \left( \Omega_\mathrm{rf} t
+ \varphi_\mathrm{mm,i} \right),\quad \textrm{i=H1,H2,V},
\end{eqnarray}
where $S_\mathrm{max,i}/2$ is the ion's fluorescence at
$\omega_\mathrm{laser}-\omega_\mathrm{ion}=-\Gamma/2$, which
corresponds to half of the maximum fluorescence for a Lorentzian
probability distribution. $S_\mathrm{mod,i}$ is the fluorescence
modulation amplitude and $\varphi_\mathrm{mm,i}$ is the phase of
the motion relative to the rf voltage. To improve the fit of the
data, the frequency of the rf trigger signal is
$\Omega_\mathrm{rf}/2$, so that the data set contains two rf
periods, i.e. $T=0...4\pi/\Omega_\mathrm{rf}$. From the
individual measurements along laser beams H1, H2 and V, the
velocity components along the trap axes are extracted using:
\begin{eqnarray}
\frac{k\, v_\mathrm{mm,y}}{\Gamma_\mathrm{nat}} & = & \frac{S_\mathrm{mod,V}\ \sqrt{1+s_\mathrm{V}}}{S_\mathrm{max,V}\ f_\mathrm{c}},\label{eq_mm_ev_1}\\
\frac{k\, v_\mathrm{mm,x}}{\Gamma_\mathrm{nat}} & = & \frac{1}{2\cos\theta_\mathrm{x}\ f_\mathrm{c}}\cdot\sqrt{A^2 + B^2 + 2AB \cos(\varphi_\mathrm{H2} - \varphi_\mathrm{H1})}\quad\textrm{and}\label{eq_mm_ev_2}\\
\frac{k\, v_\mathrm{mm,z}}{\Gamma_\mathrm{nat}} & = & \frac{1}{2\cos\theta_\mathrm{z}\ f_\mathrm{c}}\cdot\sqrt{A^2 + B^2 - 2AB \cos(\varphi_\mathrm{H2} - \varphi_\mathrm{H1})},\label{eq_mm_ev_3}\\
&& \quad \textrm{with}\quad A,B = \frac{S_\mathrm{mod,H1,H2}\cdot\sqrt{1+s_\mathrm{H1,H2}}}{S_\mathrm{max,H1,H2}}.\nonumber
\end{eqnarray}\label{eq_mm_calc}
Here, $k=2\pi/369.5\,\mathrm{nm}$ and
$\theta_\mathrm{x}=65\,^\circ$ and
$\theta_\mathrm{z}=25\,^\circ$ are the projection angles of the
laser beams H1 and H2 to the trap axes x and z, respectively.
Saturation broadening of the atomic line is taken into account by
the individual saturation parameters $s_\mathrm{i}$ of each laser beam.

In addition, the measured signal is corrected by a factor
$f_\mathrm{c}$, due to the finite lifetime of the excited state of
the ion. Only when the lifetime of the excited state is much
shorter than the modulation period of the signal
($\tau \ll T_\mathrm{rf}$), the full signal
amplitude $S^\mathrm{(0)}_\mathrm{mod}$ is observed. In general, the fluorescence of an ion,
when excited by a periodic signal, follows the differential equation $\dot{S}(t) =\tau^{-1}\, S(t) + \, S_\mathrm{drive}(t)$, 
with a periodic term $S_\mathrm{drive}(t)$~\cite{Peik_1993}. 
The general solution to this equation is:
\begin{eqnarray}\label{eq_integral}
S(t) & = & \rme^{-t/\tau} \left( c_1 + \int_{-\infty}^t S_\mathrm{drive}(t)\ \rme^{t^\prime /\tau} \rmd t^\prime \right),
\end{eqnarray}
which consists of a fast exponential decay and the damped response to the modulation.
For the fluorescence of a laser-cooled ion with micromotion, this modulation is $S_\mathrm{drive}(t) = S_0 + S_\mathrm{mod}^{(0)}\sin (\Omega_\mathrm{rf}t)$, with $S^{(0)}_\mathrm{mod} \propto |\bi{k\cdot v}_\mathrm{mm}|$. For times larger than the natural decay time $\tau$, (\ref{eq_integral}) gives:
\begin{eqnarray}
S(t)& = & \tau\, S_0 + S_\mathrm{mod}^{(0)} \cdot \frac{\tau^{-1}\sin (\Omega_\mathrm{rf}t) - \Omega_\mathrm{rf} \cos (\Omega_\mathrm{rf}t)}{\tau^{-2} + \Omega_\mathrm{rf}^2},\nonumber\\
& = & \tau\, S_0 + S^{(0)}_\mathrm{mod} \cdot \left( \tau^{-2} + \Omega_\mathrm{rf}^2 \right)^{-1/2} \cdot \sin \left( \Omega_\mathrm{rf}t+\varphi_i \right).\label{eq_solved_integral}
\end{eqnarray}
From this the modulation $m_\mathrm{det}$ of the detected fluorescence is derived to be
\begin{eqnarray}\label{eq_modulation}
m_\mathrm{det} & = & \frac{S^{(0)}_\mathrm{mod} \cdot \left( \tau^{-2} + \Omega_\mathrm{rf}^2 \right)^{-1/2}}{\tau\, S_0} = \frac{S^{(0)}_\mathrm{mod}}{S_0} \cdot f_c \, ,
\end{eqnarray}
giving a reduction in contrast compared to the modulation $m_0=S^{(0)}_\mathrm{mod}/S_0$ of
\begin{eqnarray}\label{eq_fc}
f_\mathrm{c} & = & \frac{1}{\sqrt{1+\left(\Omega_\mathrm{rf}\,\tau \right)^2}} \,.
\end{eqnarray}
With a lifetime $\tau = 8\,\mathrm{ns}$, this correction term amounts to $f_\mathrm{c}=0.61$.

The finite saturation parameter $s_\mathrm{i} \approx 0.6$ of each laser
beam leads to a reduction of the measured micromotion velocity
amplitude of about $26\,\%$. It is evaluated before the
measurement by scanning the cooling laser over the resonance and
fitting a Lorentzian profile to the scan. The fit gives the effective
linewidth and hence the saturation parameter for each laser beam
with an uncertainty of about $8\,\%$, which is the major
contribution to the systematic uncertainty of the data. Compared to
that, the uncertainty in the angles of the laser beams to the trap axes
are negligible and not taken into account.

Great care is taken to correctly subtract the background signal
due to laser straylight from the fitted off-set
$S_\mathrm{max,i}/2$. The detection laser itself contributes
about $ 5 - 10\,\%$ to the background. In the case of measuring
micromotion along V, laser beam H1 is present during the
measurement to compensate axial heating of the ion. For this
purpose the power in beam H1 is reduced by a factor of 14 and no
contribution to the modulation amplitude is detectable. Only a
constant offset is subtracted from $S_\mathrm{max,V}/2$.

The resolution achieved in the measurements is limited in general by intensity and frequency fluctuations of the spectroscopy laser. This contribution has been evaluated from repetitive measurements in all laser beams. Due to the laser beam geometry, the highest sensitivity to rf fields of $\sigma_\mathrm{rf}=21\,\mathrm{V/m}$ is reached for the micromotion y component measured along laser beam V. For the measurements along the
trap axis this uncertainty is $22\,\mathrm{V/m}$. The lowest sensitivity is obtained
in x direction with $\sigma_\mathrm{rf}=49\,\mathrm{V/m}$. The fit to the data contributes an additional  uncertainty of $8\,\mathrm{V/m}$.

As shown in figure~\ref{mm_resolution}, an electric field of
$\Delta E_\mathrm{x} = 0.9\,\mathrm{V/m}$, corresponding to $\Delta
U_\mathrm{ec}=5.8\,\mathrm{mV}$, has to be applied to produce a detectable
micromotion along the y-axis. This corresponds to an rf field amplitude of $E_\mathrm{rf,x}=(51\pm23)\,\mathrm{V/m}$, or an excess micromotion amplitude of $1.1\,\mathrm{nm}$. With 
$(\Delta\nu/\nu)_\mathrm{mm}=-v_\mathrm{mm}^2/2c^2$, this gives a relative
frequency shift due to time dilation of $-8.5\times10^{-20}$.
Here, the resolved rf field is limited only by the resolution of the dc voltage control 
on the electrodes that moves the ion radially.

\subsection{3D micromotion measurement in the trap}

\begin{figure}[hbtp]
\flushleft
\hspace{2.5cm}
\subfigure{}{\includegraphics[width=0.4\textwidth,angle=0]{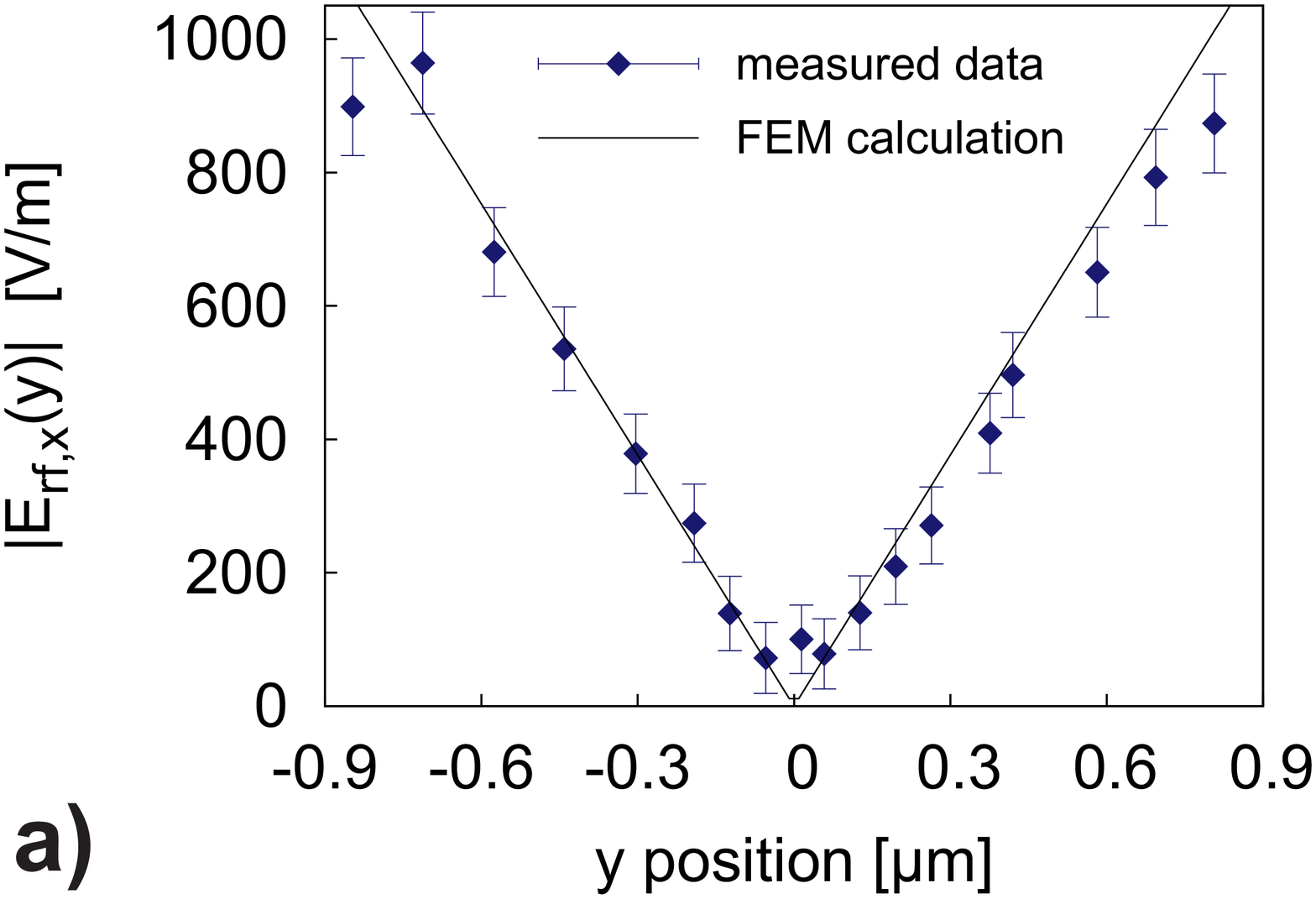}}\hspace{0.3cm}
\subfigure{}{\includegraphics[width=0.4\textwidth,angle=0]{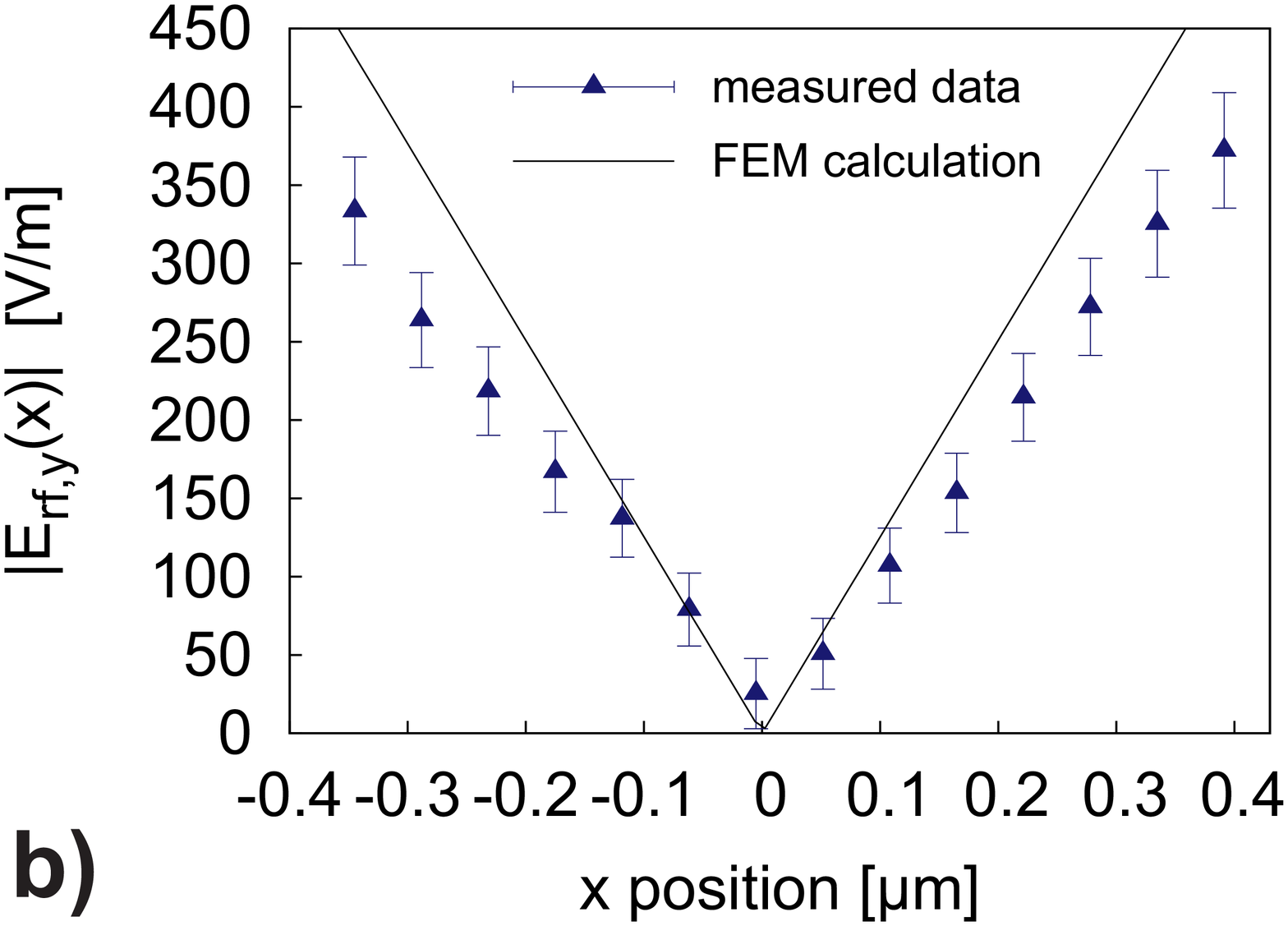}}
\caption{RF field as function of ion displacement in radial
directions x and y. To check the consistency of the measured
micromotion amplitudes, expected rf field amplitudes have been
calculated from the measured radial secular frequencies (solid
lines).}\label{radial_mm}
\end{figure}

To characterize and test the photon-correlation method
quantitatively, a single ion was first shifted in both radial
directions of the trap, where the strong rf quadrupole potential
dominates and its field gradients can be determined accurately
from our measurements of the secular frequencies. In
figure~\ref{radial_mm}a the micromotion measurement is shown for an
ion shifted along the y direction and in figure~\ref{radial_mm}b for
an ion shifted along the x direction.
The graphs show the measured rf electric field component
perpendicular to the direction of the ion shift as a function of
the ion position in the trap. The position of the ion is
calculated according to (16) in~\cite{Berkeland_1998} using the
measured radial secular frequencies and the calculated dc electric
fields applied by changing $U_\mathrm{tc}$ and
$U_\mathrm{ec}$. For comparison, the rf electric field obtained
by the FEM calculations is plotted together with the measured data
as a function of the ion position.

The measurement of the y component of the field is in good
agreement with the calculation, whereas the x component shows a
deviation of about $25\,\%$. This can be explained by the fact,
that the Rogers4350B$\texttrademark$ boards are elastic. Pictures taken from the
assembled trap stack show, that the outer compensation boards
slightly bend away from the rf trap boards in the center and
therefore have a larger distance to the trap center in x
direction. This leads to a decreased static electric field at a
given compensation voltage $U_\mathrm{ec}$. A FEM calculation,
in which the distance between the compensation and trap boards is
varied, shows a decrease of the dc field in x direction of about
$30\,\%$ for an increase of the electrode distance of $0.25\,\mathrm{mm}$, which is reasonable
looking at the taken pictures.

On the other hand, along the y direction the ion is shifted by
applying $U_\mathrm{tc}$ on the rf ground electrodes on the
quadrupole trap boards. As these electrodes are more rigidly
machined from one continuous wafer and measured secular
frequencies have confirmed the expected rf quadrupole geometry, no
relevant deviations in the applied dc fields, created by
$U_\mathrm{tc}$, are expected here. This gives a strong argument
for the good agreement between the measurement and the
calculations in the case of the vertical ion shift, where
$E_\mathrm{rf,x}$ is measured.

\begin{figure}[hbtp]
\flushleft
\hspace{2.5cm}
\includegraphics[width=0.6\textwidth,angle=0]{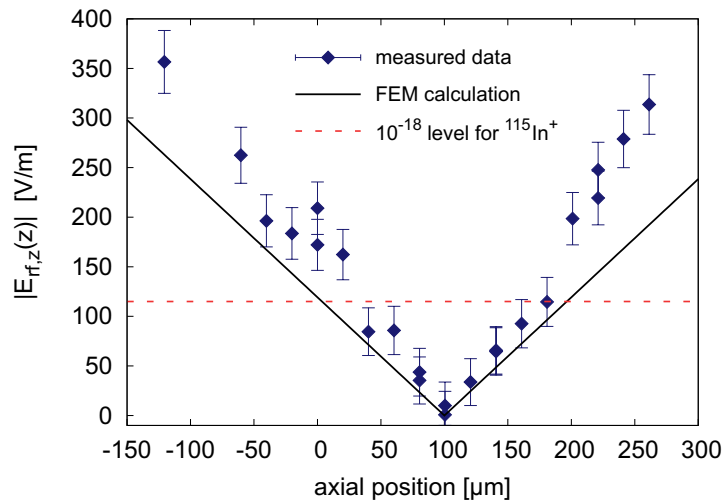}
\caption{Residual axial rf electric field along trap axis derived from measured
micromotion amplitude (diamonds) compared to axial rf field
estimated from FEM calculations (solid line). The dashed line
indicates the rf field, for which the relative frequency shift
induced by micromotion is smaller than $10^{-18}$ for an indium
ion optical clock. The zero axial position corresponds to the ion position for $U_\mathrm{ax,1}=U_\mathrm{ax,3}=4\,\mathrm{V}$, where the measurement was initiated.}\label{axial_mm}
\end{figure}

To measure residual axial rf fields, a single ion was shifted in
the direction of the trap axis and the micromotion was measured,
see figure~\ref{axial_mm}.
The laser beams have been adjusted while shifting the ion, in
order to prevent systematic uncertainties due to a change in the
intensity at the ion position. As a cross-check we measured the residual radial rf field and verified a value below $115\,\mathrm{V/m}$ in the range of $0<z<200\,\mathrm{\mu m}$.

In addition to that, FEM calculations have been carried out for comparison. The
calculations include the contribution to the rf electric field due
to the finite length of the trap as well as the slits between the
electrodes. Alignment and machining uncertainties and their effect on the
rf field have been investigated in~\cite{Herschbach_2011}. According to that, variations in the width of the slits between the electrodes due to machining tolerances would affect the measured
field gradient and can be an explanation for the slightly larger
rf fields.

Besides that, we observed a deviation of $100\,\mathrm{\mu m}$ between the ion position with minimum micromotion and the average ion position for symmetric axial dc voltages $U_\mathrm{ax,1} = U_\mathrm{ax,3} = 4.0\,\mathrm{V}$, which is at $z=0$ in figure~\ref{axial_mm}. This can be due to angular misalignment of the trap boards, 
leading to an offset in the axial rf field in the trapping region~\cite{Herschbach_2011}, machining tolerances of the electrode notches or
dc stray fields, shifting the ion from their average position in the axial trapping potential. These effects are not distinguishable in the present setup.

For optical clock operation with In$^+$ ions, the amplitudes of axial rf fields need to be 
$|E_\mathrm{rf}|< 115\,\mathrm{V/m}$
to guarantee a small enough fractional frequency shift due to time dilation of
$|(\Delta \nu/\nu)_\mathrm{mm}|=4 E_\mathrm{rf}^2 e^2/(m^2\Omega^2_\mathrm{rf}) \leq1\times10^{-18}$.
We find a length of about
$130\,\mathrm{\mu m}$ in our ion trap driven at an rf voltage
amplitude of $810\,\mathrm{V}$, within which this condition is fulfilled. At full rf amplitude of $1500\,\mathrm{V}$, this length is reduced by about a
factor of two, allowing space for about 12 trapped ions.

\section{Conclusion}

We presented a new experiment to test novel ion traps
for optical clock operation with flexible optical access to
perform measurements of micromotion in all dimensions. In our new
scalable ion trap, we are able to load and trap single
$\mathrm{^{172}Yb^+}$ ions as a high-resolution electric field
probe, as well as linear Coulomb crystals for multi-ion optical
clock operation. We detailed on the design and construction of a
novel ion trap based on stacked printed circuit boards with
on-board filter electronics and a low loss rf circuit with a
loaded quality factor of $Q_\mathrm{loaded}=640$. Using
photon-correlation spectroscopy we demonstrated the measurement of
fractional frequency shifts due to excess micromotion with a
resolution of $<10^{-19}$, which allows us to characterize trap
designs for the use of an optical clock with lowest systematic
shifts.

With this, we were able to show that already the printed circuit
board trap features a region of $130\,\mathrm{\mu m}$ along the
trap axis, in which the relative frequency shift due to
micromotion is $|(\Delta \nu/\nu)_\mathrm{mm}|
\leq1\times10^{-18}$ at a trap voltage of
$U_\mathrm{rf}=810\,\mathrm{V}$. For a multi-ion clock based on
$\mathrm{^{115}In^+}$ ions, this corresponds to a region of
$70\,\mathrm{\mu m}$ operating the ion trap at
$U_\mathrm{rf}=1500\,\mathrm{V}$ and $\omega_\mathrm{In,rad}=2 \pi
\times 1.5\,\mathrm{MHz}$, $\omega_\mathrm{In,ax}=2 \pi \times
225\,\mathrm{kHz}$. This allows to trap a chain of about 12
$\mathrm{^{115}In^+}$ ions for optical clock operation with
$|(\Delta \nu/\nu)_\mathrm{mm}|\leq10^{-18}$.

Based on the design presented and tested in this paper, an
ion trap of laser-cut AlN wafers will be machined in-house. Here,
a better mechanical stiffness and precision can be obtained and
only non-magnetic materials will be used. We expect an improvement
in the performance in terms of axial micromotion as well as heat
conductivity for precise ion trap temperature evaluation. In
addition, a larger number of trap segments will allow to increase
the number of ions used for spectroscopy.

\ack

We thank Max Harlander and Yves Colombe for exchanging experience
on low outgassing components, Christian Tamm for discussions on
detection optics and trap electronics and trap electronics and Kihwan Kim for providing us with SMD parts for preliminary tests. We thank Ekkehard Peik and Kristijan Kuhlmann for a careful reading of this manuscript. This
work was supported by the cluster of excellence QUEST.

\section*{References}

\bibliographystyle{unsrt}

\begin{thebibliography}{10}

\bibitem{Chou_2010}
Chou C W, Hume D B, Koelemeij J C J, Wineland D J and Rosenband T 2010
\newblock Frequency comparison of two high-accuracy {A}l$^+$ optical clocks
\newblock {\em Phys. Rev. Lett.} {\bf 104} 070802

\bibitem{Schmidt_2005}
Schmidt P O, Rosenband T, Langer C, Itano W M, Bergquist J C and Wineland D J 2005
\newblock Spectroscopy using quantum logic
\newblock {\em Science} {\bf 309} 749--752

\bibitem{Roberts_1997}
Roberts M, Taylor P, Barwood G P, Gill P, Klein H A and Rowley W R C 1997
\newblock Observation of an electric octupole transition in a single ion
\newblock {\em Phys. Rev. Lett.} {\bf 78} 1876--1879

\bibitem{Herschbach_2011}
Herschbach N, Pyka K, Keller J and Mehlst\"aubler T E 2011
\newblock Linear Paul trap design for an optical clock with Coulomb crystals
\newblock {\em Appl. Phys.} B DOI 10.1007/s00340-011-4790-y

\bibitem{Champenois_2010}
Champenois C, Marciante M, Pedregosa-Gutierrez J, Houssin M, Knoop M and Kajita M 2010
\newblock Ion ring in a linear multipole trap for optical frequency metrology
\newblock {\em Phys. Rev.} A {\bf 81} 043410

\bibitem{Madsen_2004}
Madsen M J, Hensinger W K, Stick D, Rabchuk J A and Monroe C 2004
\newblock {Planar ion trap geometry for microfabrication}
\newblock {\em Appl. Phys.} B {\bf 78} 639--651

\bibitem{Home_2006}
Home J P and Steane A M 2006
\newblock Electrode configurations for fast separation of trapped ions
\newblock {\em Quantum Inform. Comput.} {\bf 6} 5

\bibitem{Stick_2006}
Stick D, Hensinger W K, Olmschenk S, Madsen M J, Schwab K and Monroe C 2006
\newblock Ion trap in a semiconductor chip
\newblock {\em Nat. Phys.} {\bf 2} 36--39

\bibitem{Schulz_2006}
Schulz S, Poschinger U, Singer K and F Schmidt-Kaler F 2006
\newblock {Optimization of segmented linear Paul traps and transport of stored particles}
\newblock {\em Fortschr. Phys.} {\bf 54} 648--665

\bibitem{Hensinger_2006}
Hensinger W K, Olmschenk S, Stick D, Hucul D, Yeo M, Acton M, Deslauriers L, Rabchuk J and Monroe C 2006
\newblock T-junction ion trap array for two-dimensional ion shuttling, storage and manipulation
\newblock {\em Appl. Phys. Lett.} {\bf 88} 034101

\bibitem{Leibrandt_2009}
Leibrandt D R {\em et al} 2009
\newblock Demonstration of a scalable, multiplexed ion trap for quantum information processing
\newblock {\em Quantum Inform. Comput.} {\bf 9} 11

\bibitem{Tanaka_2009}
Tanaka U, Naka R, Iwata F, Ujimaru T, Brown K R, Chuang I L and Urabe S 2009
\newblock Design and characterization of a planar trap
\newblock {\em J. Phys.} B {\em At. Mol. Opt. Phys.} {\bf 42} 154006

\bibitem{Pearson_2006}
Pearson C E, Leibrandt D R, Bakr W S, Mallard W J, Brown K R and Chuang I L 2006
\newblock Experimental investigation of planar ion traps
\newblock {\em Phys. Rev.} A {\bf 73} 032307

\bibitem{Allcock_2011}
Allcock D T C {\em et al} 2011
\newblock Heating rate and electrode charging measurements in a scalable,
  microfabricated, surface-electrode ion trap
\newblock {\em Appl. Phys.} B DOI 10.1007/s00340-011-4788-5

\bibitem{Turchette_2000}
Turchette Q A {\em et al} 2000
\newblock Heating of trapped ions from the quantum ground state
\newblock {\em Phys. Rev.} A {\bf 61} 063418

\bibitem{Daniilidis_2011}
Daniilidis N, Narayanan S, M\"oller S A, Clark R, Lee T E, Leek P J, Wallraff A, Schulz S, Schmidt-Kaler F and H\"affner H 2011
\newblock Fabrication and heating rate study of microscopic surface electrode ion traps
\newblock {\em New J. Phys.} {\bf 13} 013032

\bibitem{Schauer_2009}
Schauer M M, Danielson J R, Nguyen A T, Wang L B, Zhao X and Torgerson J R 2009
\newblock Collisional population transfer in trapped $\textrm{Yb}^{+}$ ions
\newblock {\em Phys. Rev.} A {\bf 79} 062705

\bibitem{Fawcett_1991}
Fawcett B C and Wilson M 1991
\newblock {Computed Oscillator Strengths, and Land{\'e} g-Values, and lifetimes
  in Yb II}
\newblock {\em Atom. Data and Nucl. Data Tables} {\bf 47} 241

\bibitem{Das_2005}
Das D, Barthwal S, Banerjee A and Natarajan V 2005
\newblock Absolute frequency measurements in Yb with
  $0.08\phantom{\rule{0.3em}{0ex}}\mathrm{ppb}$ uncertainty: isotope shifts and
  hyperfine structure in the $399\textrm{-}\mathrm{nm}$
  $^{1}S_{0}\rightarrow^{1}P_{1}$ line
\newblock {\em Phys. Rev.} A {\bf 72} 032506

\bibitem{Macalpine_1959}
Macalpine W W and Schildknecht R O 1959
\newblock {Coaxial resonators with helical inner conductor}
\newblock {\em Proc. of the IRE} 2099

\bibitem{Schrama_1993}
Schrama C A, Peik E, Smith W W and Walther H 1993
\newblock {Novel miniature ion traps}
\newblock {\em Opt. Comm.} {\bf 101} 32--36

\bibitem{Berkeland_1998}
Berkeland D J, Miller J D, Bergquist J C, Itano W M and Wineland D J 1998
\newblock {Minimization of ion micromotion in a Paul trap}
\newblock {\em J. Appl. Phys.} {\bf 83} 5025--5033

\bibitem{Peik_1993}
Peik E 1993
\newblock Spektroskopie an gespeicherten In-Ionen
\newblock {\em PhD Thesis} Max-Planck-Institut f\"ur Quantenoptik, M\"unchen

\end{thebibliography}

\end{document}